\documentclass[conference]{IEEEtran}
\IEEEoverridecommandlockouts
\usepackage{cite}
\usepackage{amsmath,amssymb,amsfonts}
\usepackage{algorithmic}
\usepackage{graphicx}
\usepackage{textcomp}
\usepackage{xcolor}
\usepackage{amsthm}
\usepackage{fancyhdr}
\usepackage{booktabs}
\newcounter{note}[section]

\newtheorem{definition}[]{Definition}
\newtheorem{theorem}{Theorem}

\usepackage[ruled,vlined]{algorithm2e}

\makeatletter
\newcommand{\removelatexerror}{\let\@latex@error\@gobble}
\makeatother

\def\BibTeX{{\rm B\kern-.05em{\sc i\kern-.025em b}\kern-.08em
    T\kern-.1667em\lower.7ex\hbox{E}\kern-.125emX}}
\begin{document}

\title{Phalanx: A Practical Byzantine Ordered Consensus Protocol\\
}

\author{
\IEEEauthorblockN{1\textsuperscript{st} Guangren Wang}
\IEEEauthorblockA{\textit{Zhejiang University} \\}
\and
\IEEEauthorblockN{2\textsuperscript{nd} Liang Cai}
\IEEEauthorblockA{\textit{Zhejiang University} \\}
\and
\IEEEauthorblockN{3\textsuperscript{rd} Fangyu Gai}
\IEEEauthorblockA{\textit{University of British Columbia} \\}
\and
\IEEEauthorblockN{4\textsuperscript{th} Jianyu Niu }
\IEEEauthorblockA{\textit{Southern University of Science and Technology} \\}
}

\maketitle
\thispagestyle{fancy}
\lhead{}
\chead{}
\rhead{}
\lfoot{}
\cfoot{}
\rfoot{\thepage}
\renewcommand{\headrulewidth}{0pt}
\renewcommand{\footrulewidth}{0pt}
\pagestyle{fancy}
\rfoot{\thepage}

\begin{abstract}
Byzantine fault tolerance (BFT) consensus is a fundamental primitive for distributed computation. 
However, BFT protocols suffer from the ordering manipulation, in which an adversary can make front-running.
Several protocols are proposed to resolve the manipulation problem, but there are some limitations for them. 
The \emph{batch-based} protocols such as Themis has significant performance loss because of the use of complex algorithms to find strongly connected components (SCCs).
The \emph{timestamp-based} protocols such as Pompe have simplified the ordering phase, but they are limited on fairness that the adversary can manipulate the ordering via timestamps of transactions.
In this paper, we propose a Byzantine ordered consensus protocol called Phalanx, in which transactions are committed by \emph{anchor-based} ordering strategy.
The \emph{anchor-based} strategy makes aggregation of the Lamport logical clock of transactions on each participant and generates the final ordering without complex detection for SCCs.
Therefore, Phalanx has achieved satisfying performance and performs better in resisting ordering manipulation than \emph{timestamp-based} strategy. 
\end{abstract}

\begin{IEEEkeywords}
Blockchain, Byzantine Fault Tolerance, Ordering Manipulation, Distributed System
\end{IEEEkeywords}

\section{Introduction}
Byzantine Fault Tolerance (BFT) consensus protocols could be used to deal with arbitrary behaviors.
However, the most widely used BFT consensus protocols (PBFT\cite{castro1999practical}\cite{castro2002practical}, HotStuff~\cite{yin2019hotstuff}, etc.), limited types of malicious manners could be prevented because of the limitation of the  \emph{leader-based} system model. 
The leader always determines the next proposal to be applied, while backups can only detect whether a leader is crashed or the proposal is proposed on a duplicated serial number.
Therefore, the leader could always decide on the content of proposals and could decide the ordering of proposals to commit.
As for permissioned blockchain and some of the permissionless blockchain, they are mostly constructed based on \emph{leader-based} BFT consensus protocols that Diem (Libra) is based on DiemBFT (HotStuff~\cite{yin2019hotstuff}) and Solana takes use of protocol based on PBFT, so we need to deal with the potential risks on ordering manipulation. 

The possibilities of manipulating the ordering have left latent danger of malicious behavior. 
In recent works \cite{daian2020flash}, researchers found that the manipulation of transactions in the permissionless blockchain, such as Ethereum \cite{buterin2013ethereum}, has made the attackers grab millions of profits.
In the traditional financial system, \emph{front-running} is a well-known illegal means to grab profits, especially on Wall Street~\cite{lewis2014flash}. 
For instance, when a broker receives a market order from a customer to buy a large amount of stock, before placing the order for the customer, he or she buys a few amounts of shares of the same stock for the broker's account. 
Then, the broker places the customer's order and the price of the stock will be driven up. 
If we allow the broker to immediately sell his or her shares, a significant profit will be generated in just a short time.
The profits are just a part of the additional cost to the customer's purchase caused by the broker's self-dealing.
In the situation such as sec-killing, the priority of transactions determines whether the dealer can get the products he or she prefers.


To deal with the ordering manipulation in BFT protocols, recently, the property called \emph{order-fairness} has been widely discussed. 
Researchers have found that though collecting quorum\cite{shostak1982byzantine} votes to decide the context of transactions seems like an effective approach. However, the Condorcet Paradox \cite{brandt2016handbook} \cite{condorcet1785essay} prevents us to reach a deterministic execution ordering. 
So that, how to resolve the Condorcet Paradox is the major barrier to designing a practical protocol to achieve \emph{order-fairness}. 
Kelkar et al. proposed Aequitas\cite{kelkar2020order} and Themis\cite{kelkar2021themis}, which are \emph{batch-based} protocols and achieve the \emph{order-fairness} between by detecting strongly connected components (SCCs). 
But because of the complex graph algorithm to detect SCCs, the performance of Aequitas/Themis seems not satisfying. 
Pompe\cite{zhang2020byzantine} proposed by Zhang et al. and Wendy\cite{kursawe2020wendy} proposed by Kursawe et al. are \emph{timestamp-based} protocols.
They discard the logical clock and ordering the transactions with the timestamps.
Because of the linearity of real clock, these protocols avoid the Condorcet Paradox and simplify the ordering phase to reduce performance loss. 
However, the \emph{timestamp-based} protocols are easily destroyed by manipulating timestamps \cite{kelkar2021order}.
It might not be a good choice to construct industrial systems.
(See Section~\ref{sec:related} for details)

It should be noted that we cannot distinguish the adversarial participants from the honest ones in a distributed system, 
and recent works \cite{kelkar2020order} \cite{kelkar2021themis} \cite{zhang2020byzantine} show that we cannot make the final ordering completely independent of Byzantine nodes. 
However, we are supposed to minimize the impact of the adversary on execution ordering, 
and an effective Byzantine ordered strategy should try to eliminate malicious behavior as much as possible and reduce the impact on performance at the same time. 

In this paper, we aim to introduce a new strategy that could avoid complex graph detection and could achieve more fairness than the \emph{timestamp-based} protocols.
Based on previous work, we believe that the Byzantine ordered consensus has two properties, \emph{free will} and \emph{correctness}. 
Because of the vulnerability of real clock, we take the Lamport Logical Clock as the indicator to order transactions. 
Then, to achieve the fairness, we make aggregation of each participant preferences on transactions ordering to find the \emph{anchor commands} with linear reliable context to make a commitment. 
It has prevented the occurrence the Condorcet Paradox, and the complex graph detection has been avoided. 
So that, we design, implement, and evaluate Phalanx, a \emph{anchor-based} Byzantine ordered consensus protocol.
It avoids complex graph detection and reflects the real preference of each node to commit transactions. 
In summary, we make the following contributions:
\begin{itemize}
    \item We propose \emph{anchor-based} ordering strategy, which has avoided graph detection and could resist ordering manipulation. 
    \item We propose a protocol called Phalanx which has achieved \emph{anchor-based} ordering and we implement it as a modular component. We release it for public use\footnote{https://github.com/Grivn/phalanx}.
    \item We conduct a comprehensive evaluation of Phalanx for performance and reliability. Phalanx does not introduce too much performance loss and has advantages in the WAN environment.
    And Phalanx performs better in resisting ordering manipulation than \emph{timestamp-based} strategy.  
\end{itemize}

\section{System Model and Anchor-Based Ordering}

In this section, we firstly present the system model, then define ordering properties.
Finally we introduce our strategy to prevent ordering manipulation with logical clock, and how to avoid the occurrence of the Condorcet Paradox. 

\subsection{System Model} 
We consider a system of $n = 3f + 1$ consensus nodes (hereinafter referred to as nodes for short) and any number of proposers (or we can call them clients). 
In this paper, we focus on the adversary in nodes and ignore the potential malicious behavior of proposers.
We assume that an adversary can control up to $f$ nodes. 

\noindent\textbf{Command and Context.} 
The \emph{command} is the atomic unit for our Byzantine ordered consensus system. 
It is generated by proposers. 
Let $r$ denotes command. For commands $r_1$ and $r_2$, we use notion $\prec$ to describe their context, which means the sequential relationship between $r_1$ and $r_2$.
If $r_1$ should be committed before $r_2$, then $r_1 \prec r_2$.
The command can be assigned with sequence number to indicate the preference ordering for the proposer.
Whenever proposer $P_i$ is trying to propose request $m$, it should generate a command with data structure $\langle i,n,d,m \rangle$, in which 
$i$ is the identifier of proposer, 
$m$ is the request content, 
$n$ is a monotonically increasing sequence number which indicates the sending order for current command on $P_i$, 
$d$ is the digest of command that $d \leftarrow h(\langle i,n,m \rangle)$.

\noindent\textbf{Partial Ordering and Total Ordering.}
The commands generated by proposers would be broadcast to every node.
The node $N_i$ needs to generate its \emph{partial ordering} according to the order of reception.
And $N_i$ generates logs to describe its partial ordering.
Each node has a logical clock starting from 0.
Every time $N_i$ has received command $r$, the logical clock is advanced by 1.
Then, $N_i$ generates a \emph{log} for $r$ assigned with current logical clock. 
The structure is $\langle i,n,d_r \rangle$, in which 
$i$ is the identifier of $N_i$, 
$n$ is the logical clock stamp when current \emph{log} has been generated, 
$d_r$ is the digest of $r$ that $d_r \leftarrow r.d$.

Taking the \emph{partial ordering} of each node as the material, Byzantine ordered consensus would make all the trusted nodes obtain a consistent \emph{total ordering}.
The \emph{commands} would be committed according to the consistent \emph{total ordering} and would send back reply message to specific proposers.

Let $o$ denotes \emph{log}. Call that $o \to r$, if and only if $o.d_r=r.d$. 
Let $\prec_i$ denote the context in $N_i$'s partial ordering. 
For logs $o_1$ and $o_2$ from $N_i$, there are $r_1$ and $r_2$ that $o_1 \to r_1$ and $o_2 \to r_2$. 
There is a context that $r_1 \prec_i r_2$ if and only if $o_1.n<o_2.n$.
Although the partial ordering for each node might- be different,
every non-faulty node would eventually find the same total ordering to commit commands.

\subsection{Network Model}
The modular implemented Phalanx component should be combined with a BFT protocol so that the network assumptions should support both Phalanx component itself and the combined original BFT protocol.

The communication in Phalanx component is asynchronous, so that, in addition to the network assumption of the original BFT protocol, we only need to satisfy the messages sent from one participant will reach the recipient(s) eventually. As asynchronous network assumption is the weakest one in BFT protocol groups' network assumptions, the Phalanx component will not impact the robustness of the original BFT cluster.

Besides, there is another basic assumption called \emph{sending-receiving} model that messages receiving order should always be the same as their sending order. For instance, $N_1$ is trying to send messages toward $N_2$. If $N_1$ has sent $m_1$ before $m_2$, $N_2$ must receive $m_1$ before $m_2$. It could be achieved easily with the use of sequence numbers.

\subsection{Ordering Properties}

In state-of-the-art leader-based BFT consensus protocols, the total ordering is determined by the leader and followers can only detect limited malicious behaviors. 
It doesn't respect the collective preferences of each correct participant, 
and, because a leader has controlled the total ordering, there is a risk of suffering a manipulation attack, which means the adversaries make the transactions more likely to be committed according to their own wishes.


Byzantine ordered consensus protocol is used to tolerate \emph{Byzantine faults} in ordering manipulation. 
To achieve it, we believe Byzantine ordered consensus protocol should have two essential properties, free will and fairness. 

\noindent\textbf{Free Will}. 
Free will is a concept proposed by Zhang et al. \cite{zhang2020byzantine}.
Briefly, it means no one could manipulate the total ordering directly. 
For instance, here is a pair of commands $r_1$ and $r_2$.
If their context in each node's partial ordering indicates $r_1 \prec r_2$ (or $r_2 \prec r_1$), their context in the total ordering must be the same that $r_1 \prec r_2$ (or $r_2 \prec r_1$). 
If some of the nodes prefer $r_1 \prec r_2$ while others $r_2 \prec r_1$, in the total ordering of trusted node, the context between such a pair of commands should be determined by every participant's preference which means both of them have an opportunity to be committed at first.


\noindent\textbf{Fairness.}
The ordering in reality deserves some intuitive expectations as below.
1) The commands proposed by the same proposer may born with context that for $r_1.i=r_2.i$, we have $r_1 \prec r_2$, iff $r_1.n<r_2.n$. 
2) The commands proposed by different proposers may deserve context because of the physical limitations. For instance, $P_1$ and $P_2$ are located in the same place. If $P_1$ has proposed $r_1$ several seconds earlier than $P_2$ proposing $r_2$, $r_1$ deserves the context $r_1 \prec r_2$ in the total ordering commitment.
Although, the ordering can be affected by multiple factors that we cannot find the totally correct ordering, we can make the context in total ordering show the intuitive expectations to the maximum. 

By using some reasonable ordering strategies, which take every node's partial ordering into thought, we can always achieve certain free-will characteristics.
To achieve better fairness, we need to aggregate the logical clock from each participant. 
However, the Condorcet Paradox prevents us to find distinct total ordering.
Next, we would like to introduce our method to deal with the paradox.

\subsection{Anchor-Based Ordering}

\begin{definition}[Reliable Context]
    For a pair of commands $r_1$ and $r_2$, there is a reliable context $r_1 \prec r_2$ (or $r_2 \prec r_1$), iff at least one non-faulty node $N_i$ believes $r_1 \prec_i r_2$ (or $r_2 \prec_i r_1$).
\end{definition}

\begin{theorem}[]
    Let $\prec_r$ denote the reliable context. For commands $r_1$ and $r_2$, if there are $f+1$ nodes believe $r_1 \prec r_2$, then there is reliable context $r_1 \prec_r r_2$.
\end{theorem}

For the reliable context is generated by the non-faulty nodes, selecting it into the total ordering satisfy the \emph{free-will}. If the context is not reliable, it means the context could be generated by a byzantine node, and might be the will of the adversaries. It is risky to select such a unreliable context into total ordering.

\noindent\textbf{Linear Anchor.}
To prevent the Condorcet Paradox, we need to find something with linearity. 
Here, a series of commands with linear reliable context can help us commit the commands. 
We call such series of commands as \emph{anchor commands}. 
Let $r_a$ denote a \emph{anchor command}. 
With the reliable context between anchor commands, we find a partial ordering set $\langle \{r_a\}, \prec_r \rangle$.

Then we need to commit the anchor commands according to the partial ordering set.
Each time when we commit an anchor command $r_a$, 
we need to generate command set $F$ which is constructed with $r_a$ and the commands $\{r'\}$ with reliable context $r' \prec_r r_a$.
After that, the commands would be committed linearly, which has avoided the Condorcet Paradox, and make the ordering process more practical and conducive to implementation.

In the next section, we would like to introduce the protocol we design with the \emph{anchor-based} ordering strategy.

\section{Phalanx}

Phalanx is a Byzantine ordered consensus protocol that can be implemented as a modular component and combined with any state-of-the-art BFT consensus algorithms, without neither additional hardware support nor additional network assumptions. 
It could help the BFT protocols to achieve \emph{anchor-base} ordering strategy, and not take much effect on the original protocol. In this section, we would like to introduce Phalanx and make proof of its basic properties.

\subsection{Overview}

Phalanx has two types of roles: proposers and nodes. 
A proposer $P_i$ could propose command $c$ with given order. 
The non-faulty nodes $\{N_i\}$ are trying to reach consensus on the total ordering of commands. 

\begin{figure}[ht]
\centering
\includegraphics[scale=0.55]{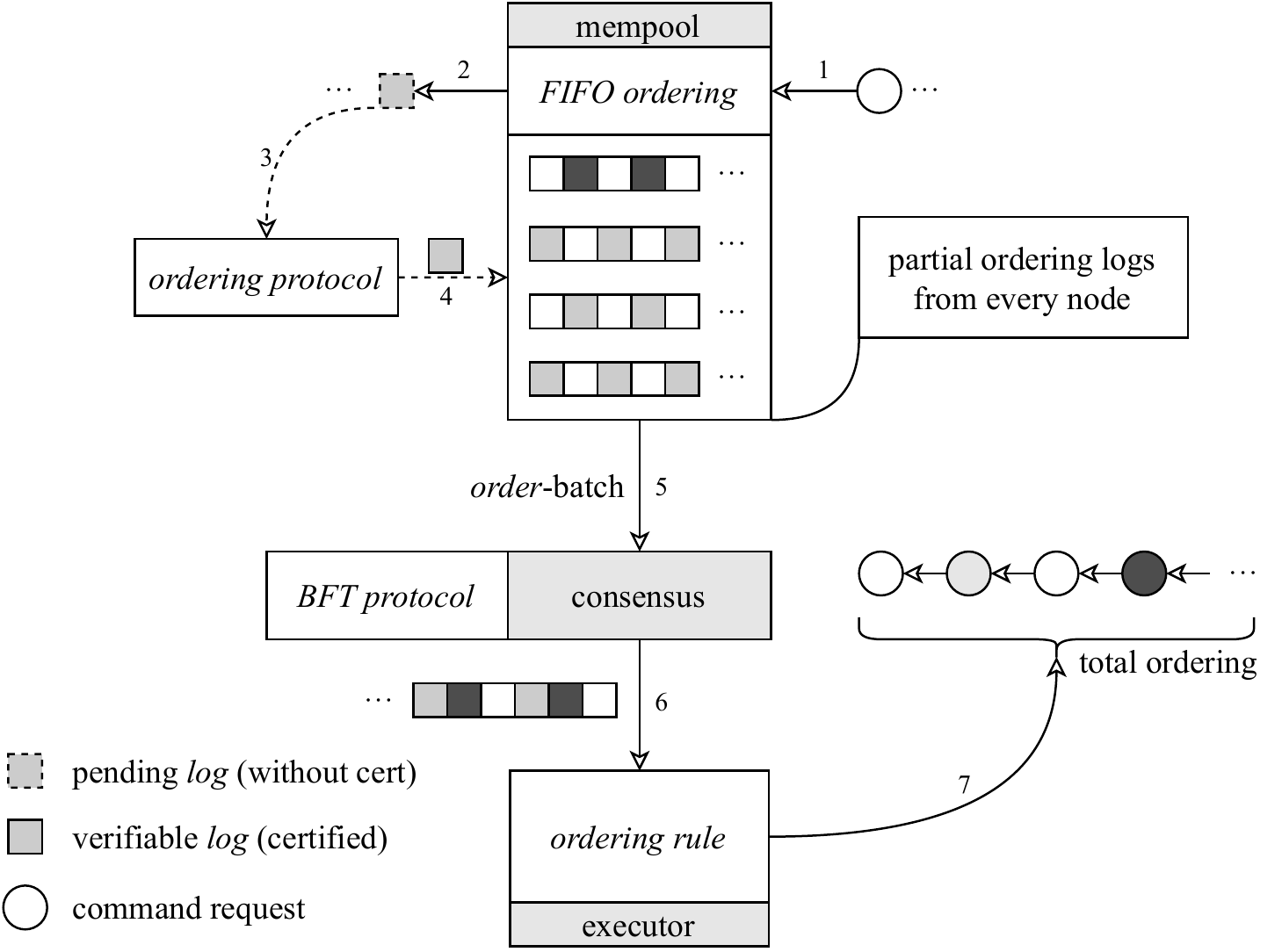}
\caption{The workflow for Phalanx nodes.}
\label{fig:workflow-node}
\end{figure}

As is shown in Fig. \ref{fig:workflow-node}, a node is constructed with three main parts: \emph{mempool}, consensus, and \emph{executor}.
Whenever node $N_i$ has received the command sent from proposers, firstly, put them into $N_i$'s partial ordering according to FIFO (First-In-First-Out) receiving the ordering. 
Then, pending \emph{log}s for $N_i$ are created. 
After the process of \emph{order-protocol}, verifiable \emph{log}s are generated to notify others its own partial ordering. 
The \emph{mempool} is a container of \emph{log}s received from each node. 
In consensus module, using the \emph{log}s in \emph{mempool}, we generate \emph{order}-batch to trigger consensus and get the same \emph{log} stream for non-faulty nodes with the help of any state-of-the-art BFT protocols. 
Taking the \emph{log} stream as material, in \emph{executor}, the non-faulty nodes eventually find the same total ordering for commands.
In particular, we list some basic notations as needed in Table~\ref{table:Notation}.
As for the sets and vectors, unless otherwise specified, they do not have preset size.
If the vector have preset size, every element in it should have a initial value.

\begin{table}[t]
    \centering
    \caption{\textbf{Basic Notations.}}
    \begin{tabular}{ l l }
        \toprule[1pt]
        Notation                     & Description \\
        \midrule

        $r$                          & Command. \\
        $\prec$                      & Context. \\
        $\prec_r$                    & Reliable context. \\
        \midrule
        
        $N$                          & Node. \\
        $N_i$                        & The node with identifier $i$. \\
        $P$                          & Proposer. \\
        $P_i$                        & The proposer with identifier $i$. \\
        \midrule
        
        $\bot$                       & Empty value. \\
        $A$                          & A set or a vector \\
        $A[i]$                       & The element in $A$ whose keyword is $i$. \\
        $\mathbb{A}$                 & FIFO queue. \\
        $\mathbb{A}.front()$         & Take out the front-\emph{log} in the FIFO queue. \\
        $\mathbb{A}.read\_front()$   & Read the front-\emph{log} without taking it out. \\
        $\mathbb{A}.remove\_front()$ & Remove the front-\emph{log} in the FIFO queue. \\
        $\mathbb{A}.push\_back(v)$   & Add the element $v$ to the end of queue. \\
        $\mathbb{A}.len()$           & The number of elements in current queue. \\
        \midrule

        $h(m)$                       & Calculate the hash of $m$. \\
        $\rho_i$                     & Partial signature generated with private key. \\
        $psig_i(e)$                  & Generate partial signature for $e$. \\
        $pverify(\rho_i)$            & Verify the partial signature $\rho_i$. \\
        $agg(e,\{\rho_i\}_{i\in P})$ & Aggregate the partial signatures $\{\rho_i\}_{i\in P}$ for $e$. \\
        $verify(sig)$                & Verify the aggregated signature with public key. \\
        
        \bottomrule[1pt]
    \end{tabular}
    \label{table:Notation}
\end{table}

\subsection{Cryptographic Notations}
We use standard hash function (e.g., SHA256) to generate the digest of messages. 
Generate the digest of $m$, $d\leftarrow h(m)$, then $d$ could be regarded as the identifier of $m$. 
If there is another message $m'$, we say that $m'$ is the same as $m$ if and only if $h(m')=h(m)$.

We assume a public key infrastructure (PKI) of each participant in our system including proposers and nodes which means each of them is identified by its own public key. 
When proposer $P_i$ or node $N_i$ is trying to send message $m$, 
the communication transmission will be carried out reliably through $\langle m\rangle_i$. 

Besides, we make use of ($k,n$)-threshold signature among nodes.
There is a single public key and each $N_i$ holds a distinct private key. 
If event $e$ needs to be determined by the vote of the nodes.
Sending a vote means approval, and not sending a vote means denial.
Each time when $N_i$ sends a vote, it generates a partial signature with its private key $\rho_i \leftarrow psig_i(e)$.
We could verify the validation of partial signature with $\rho_i$ with $pverify(\rho_i)$. 
When there is a set of valid partial signatures $\{\rho_i\}_{i\in P}$ and $|P|=k$, generate aggregated signature for $e$, $sig\leftarrow agg(e,\{\rho_i\}_{i\in P})$. 
Every node could verify the aggregated signature with the single public key that $verify(sig)$. 
If it is a valid aggregated signature, the function should return $true$ value.

The aggregated signature could be regarded as a kind of \emph{certificate}, denoted as $cert$.
If $cert$ is valid, then it means there are $k$ nodes have voted on event $e$.
Make use of $(2f+1,n)$-threshold signature to generate $cert$ for $e$ in BFT consensus algorithm, and we could verify the $cert$ to check the quorum agreement on $e$.

\subsection{Mempool}
Each node $N_i$ has its own mempool. 
Let $\mathrm{mempool}_i$ denote it. 
Whenever $N_i$ receives commands from proposers, the $\mathrm{mempool}_i$ would receive them and generate $N_i$ partial ordering according to FIFO receiving ordering. 
At the same time, certified \emph{log}s would be generated with \emph{ordering protocol} and notify other participants. 

\begin{table}[t]
    \centering
    \caption{\textbf{Structures Notations for Mempool.}}
    \begin{tabular}{ l l }
        \toprule[1pt]
        Notation        & Description \\
        \midrule

        $\mathrm{mempool}_i$ & The mempool module for node $N_i$. \\
        \midrule

        $o$                          & The verifiable \emph{log} to describe ordering preference. \\
        \bottomrule[1pt]
    \end{tabular}
    \label{table:StructuresNotationMempool}
\end{table}

\vspace{1mm} \noindent \textbf{Structures.}
The structure notion for mempool is shown in Table~\ref{table:StructuresNotationMempool}.
To make the \emph{log} verifiable, we upgrade the structure of $o$ in section 3. 
The verifiable $o$ is $\langle i,n,t,d_r,d_{pre},d_{cur},cert \rangle$, in which 
$t$ is the timestamp when we generate current \emph{log}, 
$d_{pre}$ is the digest of previous \emph{log}, which refers to the \emph{log} with sequence number $n-1$ (if $n$ is equal to 1, then $d_{pre}$ is empty), 
$d_{cur}$ is the digest of current \emph{log} that $d_{cur} \leftarrow h(\langle i,n,t,d_r,d_{pre} \rangle)$,
$cert$ is a certificate generated by \emph{ordering protocol} with $2f+1$ nodes and it could be used to verify the validation of current \emph{log}.

\begin{table}[t]
    \centering
    \caption{\textbf{States Notations for Mempool.}}
    \begin{tabular}{ l l }
        \toprule[1pt]
        Notation        & Description \\
        \midrule
        $seq$           & An integer, initialized to 0, is used to track the latest \emph{log}  \\
                        & generated by $N_i$ that it is $N_i$ logical clock. \\
        \midrule
        
        $o_{pending}$   & A \emph{log} type element, initialized to $\bot$, which is used to track \\ 
                        & the \emph{log} without certificate. \\
        \midrule

        $H$             & An $n$-sized vector of \emph{log}, where $H[j]$, initialized to $\bot$,  \\
                        & is used to track the latest verifiable \emph{log} from $N_j$. \\
        \midrule

        $\mathbb{R}_F$  & A FIFO queue command, is used to receive the commands \\
                        & from proposers. \\
        
        \bottomrule[1pt]
    \end{tabular}
    \label{table:StatesNotationMempool}
\end{table}

\vspace{1mm} \noindent \textbf{States.}
To operate the \emph{ordering protocol}, $N_i$ needs to maintain local states as is shown in Table~\ref{table:StatesNotationMempool}. The $seq$ is used to track the latest \emph{log}. The $o_{pending}$ is used to track the \emph{log} without certificate. The $n$-sized vector $H$ is used to track the latest verifiable \emph{log}. The $\mathbb{R}_F$ is used to receive the commands from proposers.

\begin{figure}[ht]
\centering
\includegraphics[scale=0.6]{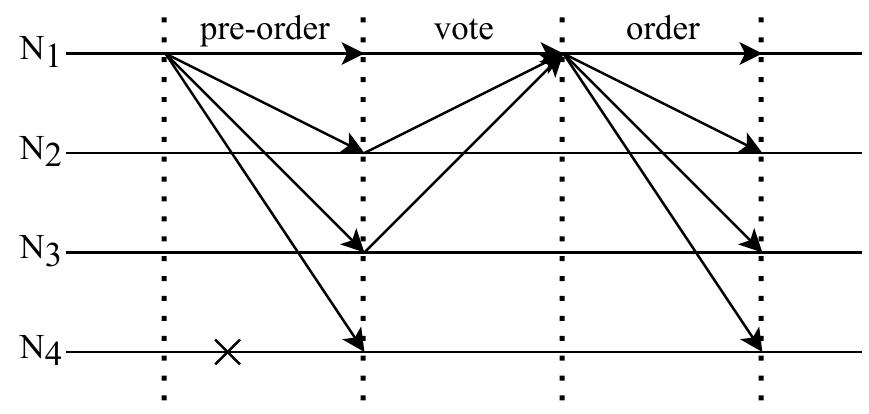}
\caption{
Here are 4 nodes and $N_4$ is crashed. 
When $N_1$ is trying to generate verifiable \emph{log}s, $N_1$ should broadcast pre-order message at first. 
Then, each node that received pre-order message should verify it and send back vote to $N_1$ if it is legal. After $N_1$ has collected quorum (here is 3) votes from every participant including itself, generate a verifiable \emph{log} and broadcast it.}
\label{fig:ordering-protocol}
\end{figure}

\vspace{1mm} \noindent  \textbf{Protocol.} As is shown in Fig. \ref{fig:ordering-protocol}, nodes run \emph{order-protocol} to generate verifiable \emph{log}s. The steps are as follows.

Step 1: Pre-Order. 
If the FIFO queue $\mathbb{R}_F$ isn't empty and there isn't a pending \emph{log} that $o_{pending}=\bot$, $N_i$ would take the front element $r \gets \mathbb{R}_F.front()$ and then try to generate verifiable \emph{log} for it. $N_i$ advances its local logical clock $seq \gets seq + 1$, and generates the order tuple without certificate $o \gets \langle i,n,d_c,d_{pre},d_{cur} \rangle$, $o.n \gets seq$, $o.d_c \gets c.d$, $o.d_{pre} \gets H[i].d_{cur}(o.n>1)$. 
Store the tuple $o$ as a pending \emph{log} that $o_{pending} \gets o$. Then, $N_i$ broadcasts pre-order message $\langle$PRE\_ORDER $o \rangle_i$ and starts to wait for the responses from $2f+1$ nodes (including itself).

Step 2: Vote. 
A node $N_j$ (including $N_i$ itself) receives the pre-order message sent from $N_i$ and tries to verify the tuple $o$ with $N_i$'s latest verifiable \emph{log} that $o_{h} \gets H[i]$. 
First, $o$ should have a valid digest. 
Next, if $o_h$ is $\bot$, then $o.n$ should be equal to 1. 
If $o_h$ is not $\bot$, then $o$ should be the next log of $o_h$ that $o.n=o_h.n+1$ and $o.d_{pre}=o_h.d_{cur}$. 
If $o$ satisfies these requirements above, it could be regarded as a valid tuple. 
Then, the node $N_j$ would respond with $\langle$VOTE $d,\rho_j \rangle_j$, where $d$ is the digest of $o$ that $d \gets o.d_{cur}$, and $\rho_j$ is a partial signature generated by $N_j$ for digest $d$ that $\rho_j \gets psig_j(d)$. Whenever $N_j$ has voted \emph{log} $o$ from $N_i$, $N_j$ will not vote for another \emph{log} $o'$ from $N_i$ that $o'.n=o.n$.

Step 3: Order. 
Whenever $N_i$ receives response $v_j$ from $N_j$, it verifies the partial signature with $pverify(v_j.\rho_j)$. 
If it is valid, accept it. 
When $N_i$ has received valid responses from $2f+1$ nodes (including itself), there is a set of valid partial signatures $\{\rho_j\}$ and $|\{\rho_j\}|$ is equal to $2f+1$. 
Generate aggregated signature with the set and complete \emph{log} that $o.cert \gets agg(o.d_{cur}, \{\rho_j\})$. 
Next, $N_i$ broadcasts partial order message $\langle$ORDER $o \rangle_i$ to notify others with its latest verifiable partial order. 
When node $N_j$ receives the \emph{log} $o$ from $N_i$, verify it with $verify(o.cert)$. 
If it's valid, then update the latest \emph{log} of $N_i$ in local states that $H[i] \gets o$.

\begin{figure}[ht]
\centering
\includegraphics[scale=0.65]{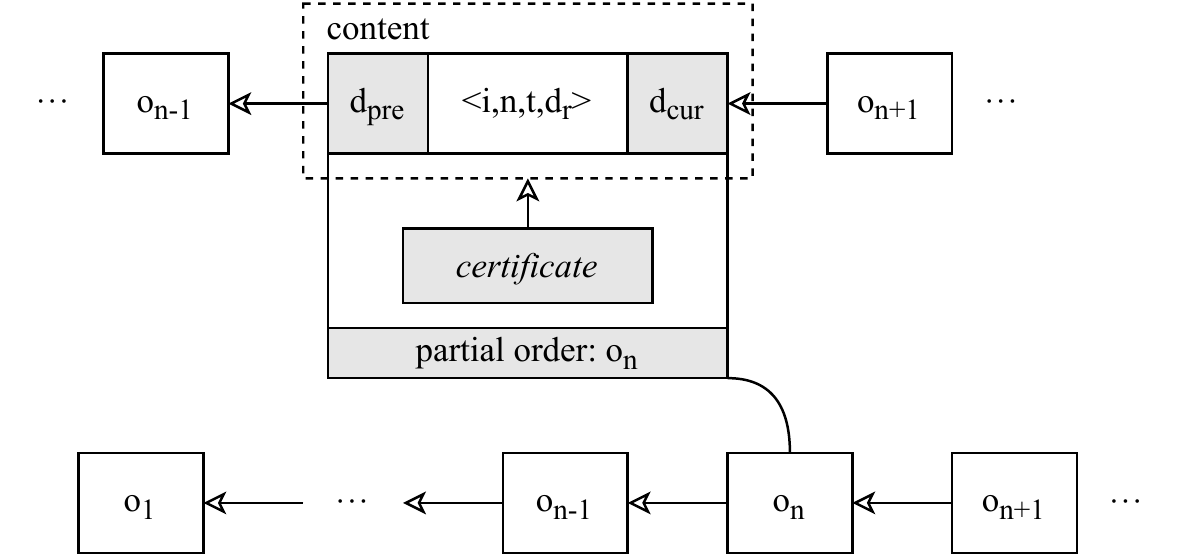}
\caption{The verifiable \emph{log}s generated by the same node form a chain structure with hash digest.}
\label{fig:ordering-logs}
\end{figure}

At last, the \emph{log}s generated by $N_i$ is constructed like Fig. \ref{fig:ordering-logs}. 
The subscript of $o_k$ in the figure indicates the \emph{log}'s serial number. 
The digests of \emph{log}s construct a chained structure and the certificate could be used to verify the validation of each \emph{log}.
Besides, $\mathrm{mempool}_i$ stores the commands and verifiable \emph{log}s $N_i$ has received. 
Every time trying to use them, we could get the command $r$ according to its digest and get \emph{log} $o$ according to tuple $\langle i,n \rangle$, where $i$ is the author of partial order and $n$ is the sequence number.

\subsection{Consensus}

Let $\mathrm{consensus}_i$ denotes the consensus module for node $N_i$. It is used to find the same \emph{log} stream for each non-faulty node. 
To achieve this goal, Phalanx could employs any state-of-the-art BFT protocols, such as PBFT, HotStuff, and even HoneyBadgerBFT, an asynchronous protocol. 
The primitive BFT protocol could help the non-faulty node reach consensus on the order of $tx$-batch. Here, in Phalanx, the contents of consensus batch are replaced by the latest \emph{log}s for each node. 
We call it \emph{order}-batch. 
After the consensus process of primitive BFT protocol, we can find the consistent \emph{log}s set to generate total ordering in the executor module.

\begin{table}[t]
    \centering
    \caption{\textbf{Structure Notations for Consensus.}}
    \begin{tabular}{ l l }
        \toprule[1pt]
        Notation        & Description \\
        \midrule
        
        $\mathrm{consensus}_i$ & The consensus module for node $N_i$. \\
        \midrule
        
        $b$             & An \emph{order}-batch, which is constructed with element $H$. \\
        \midrule
        
        $H$             & An $n$-sized vector of \emph{log}, where $H[i]$, initialized to $\bot$,  \\
                        & stores the latest verifiable \emph{log} from $N_i$. \\
        \midrule
        
        $S$             & A set of \emph{log}s $\{o\}$, in which the elements are the \emph{log}s \\
                        & from every node. \\
        \bottomrule[1pt]
    \end{tabular}
    \label{table:StructuresNotationConsensus}
\end{table}

\vspace{1mm} \noindent  \textbf{Structures}. 
The consensus module takes use of some basic data structures as is shown in Table~\ref{table:StructuresNotationConsensus}. 
We take the \emph{order}-batch $b$ into the primitive BFT consensus process. 
After the consensus agreement of the ordering to submit $b$, non-faulty nodes can generate the same \emph{log}s set $S$.
In addition, the elements in $S$ should be sorted by the strategy that 
sort the \emph{log}s in ascending order according to the sequence number $o.n$, 
and if \emph{log}s have the same sequence number, 
sort them in ascending order according to the generators' identifier $o.i$.

\begin{table}[t]
    \centering
    \caption{\textbf{States Notations for Consensus.}}
    \begin{tabular}{ l l }
        \toprule[1pt]
        Notation        & Description \\
        \midrule
        $V$             & An $n$-sized vector of integers where $V[i]$, initialized with 0, \\
                        & tracks the sequence number of \emph{log}s which is committed in \\
                        & consensus module from $N_i$, and we could regard $V$ as \\
                        & a kind of vector clock \\
        \midrule
        
        $\mathbb{S}$    & A FIFO queue for sets of \emph{log}s $S$. \\
        \bottomrule[1pt]
    \end{tabular}
    \label{table:StatesNotationConsensus}
\end{table}

\vspace{1mm} \noindent  \textbf{States}. 
Each node $N_i$'s consenter maintains the basic local states as is shown in Table~\ref{table:StatesNotationConsensus}. The $n$-sized verctor $V$ is used to track the sequence number of committed \emph{log}s. The FIFO queue $\mathbb{S}$ is used to store the \emph{log}s. 

\vspace{1mm} \noindent \textbf{Actions}. If we employ leader-based BFT protocol, such as PBFT or HotStuff, the \emph{order}-batch could be always generated by the leader in the current view (or round), so that we could provide a standard interface to generate \emph{order}-batch. 
Whenever the leader has found a latest \emph{log} for $N_i$ in mempool which has a larger sequence number than $V[i]$, it could try to generate \emph{order}-batch $b$ and trigger the BFT protocol. 
The content in $b$ could be generated from mempool that $b.H \gets \mathrm{mempool}_i.H$. 
After the process of BFT protocol, every non-faulty node will find $\{b_1, b_2, ..., b_m,...\}$, in which there are a series of \emph{order}-batches in the same order.

Next, commit the \emph{order}-batches in $\{b_m\}$ one by one. Whenever trying to commit $b_m$, a set of \emph{log}s $S$ would be generated. The steps are as follows.

Step 1. Verify the certificates of \emph{log}s in $b_m.H$. If there is an invalid \emph{log}, stop the process and change the leader in BFT protocol. If they are valid, initialize an empty \emph{log} set $S$. For each \emph{log}s in $b_m.H$, we should process it according to step 2.

Step 2. Let $n_h \gets b_m.H[j].n$ and $n_c \gets V[j]$. If $n_h \leq n_c$, just return. If $n_h>n_c$, then (1) update $V[i]$ with $n_h$, (2) add $b_m.H[j]$ into $S$, (3) get the \emph{log}s generated by $N_j$ with sequence number belonging to $(n_c,n_h)$ from mempool and add them into $S$. If we cannot find the expected \emph{log}s from mempool, request other nodes for it and verify the digests and certificates.

Step 3. After the process of each \emph{log} in $b_m.H$, add $S$ into the end of $\mathbb{S}$.

As for employing leadless BFT protocols, the generation of \emph{order}-batch depends on the generation strategy of the primitive protocol. For instance, in HoneyBadger-BFT (HB-BFT), each node would generate a slice of the proposal, and the final committed block is constructed by part of them. So that, if we employ Phalanx in HB-BFT, node $N_i$ would just propose its latest \emph{log} to trigger BFT protocol. After each time the BFT protocol consensus process is finished, an \emph{order}-batch could be constructed. At last, we could find a series of \emph{order}-batches in the same order and commit them one by one to generate the \emph{log} sets to update $\mathbb{S}$.

After the process of consensus module, each non-faulty node would find the same FIFO queue $\mathbb{S}$. The FIFO queue $\mathbb{S}$ is going to be used in \emph{executor} module to generate total ordering.

\subsection{Executor}

\begin{figure*}[ht]
\centering
\includegraphics[scale=0.45]{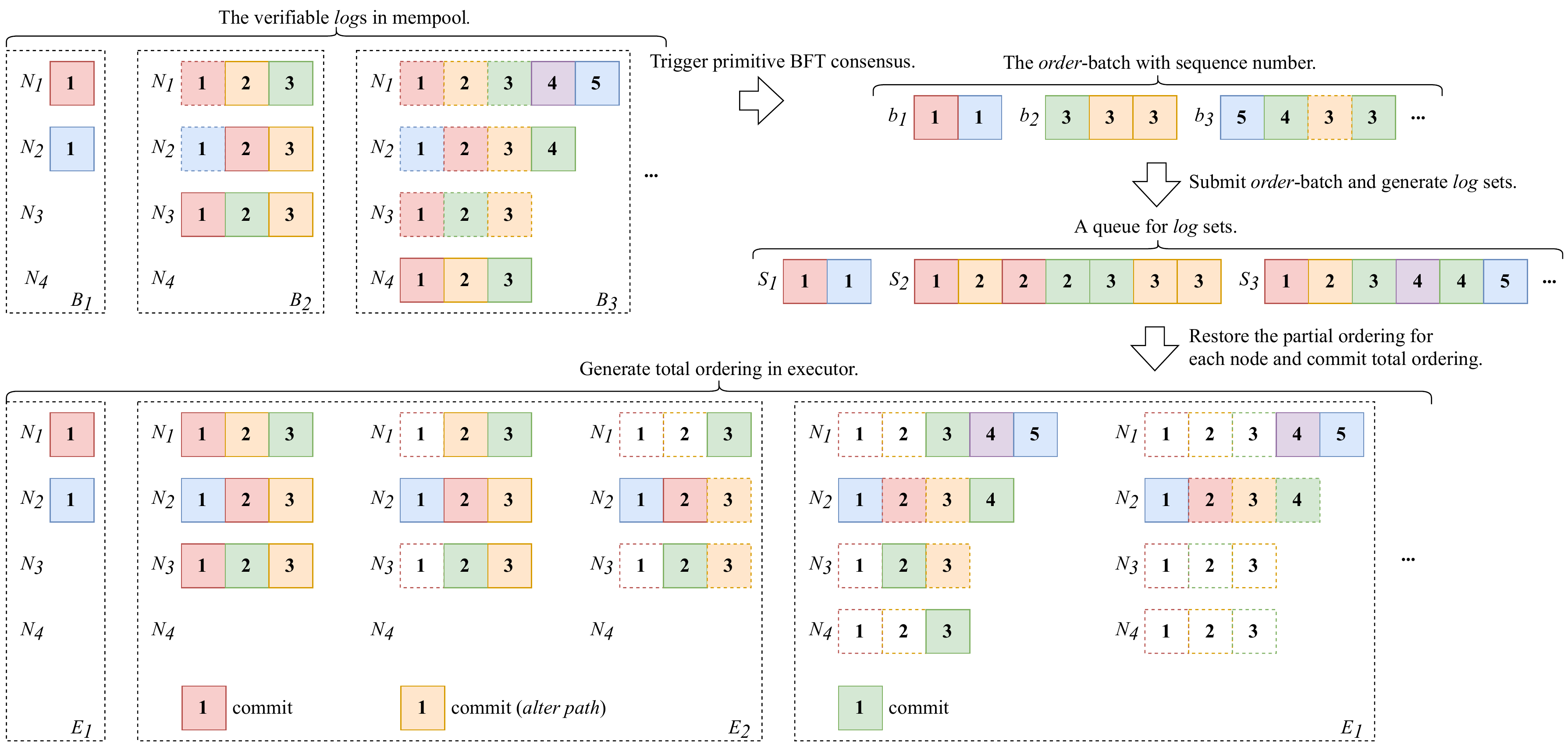}
\caption{
This figure indicates the process Phalanx to generate total ordering in a 4 nodes cluster.
Assume that the primitive BFT protocol here is a leader-based protocol and the leader is $N_1$. 
The cubes have the same color denote the verifiable \emph{log}s pointing to the same command. 
With the help of protocol in mempool, each node has generated its own partial ordering and the verifiable \emph{log}s are stored in mempool. 
The first time $N_1$ trying to trigger primitive BFT consensus, it has only received verifiable \emph{log}s from $N_1$ and $N_2$ as is shown in $B_1$.
Then, generate \emph{order}-batch $b_1$. 
Next, while trying to generate the second batch, the \emph{log}s in \emph{mempool} is shown as $B_2$, so that generate \emph{order}-batch $b_2$.
The third \emph{order}-batch is constructed as $b_3$ while the \emph{mempool} shown as $B_3$.
The \emph{order}-batches will be confirmed on non-faulty nodes via BFT primitive consensus one by one. 
The non-faulty node should submit $b_1$, $b_2$, and $b_3$ by order, and generate the corresponding \emph{log} sets $S_1$, $S_2$, and $S_3$. 
In executor, we need to restore each node's partial ordering, and then try to commit the commands. 
For the following description, the dashed cubes indicate the commands which have already been committed, and the blank cubes indicate the corresponding element in the queue has been removed. 
With the help of FIFO queue, the \emph{log} sets will be submitted into executor by order.
After we submit $S_1$, the FIFO queues in \emph{executor} is constructed as $E_1$. 
Now, we cannot find any anchor command.
After we submit $S_2$, the FIFO queue in \emph{executor} is shown as $E_2$. 
Then, we find $N_1$ and $N_3$ believe the red command should be committed at first, so that the red command becomes the anchor command. 
Commit the red command. 
After the commitment of red command, in $E_2$, the yellow command be selected as anchor command via \emph{alter path} and the anchor-set contains the yellow command only. 
Then, commit the yellow one.
After we submit the \emph{log} set $S_3$, the procession in \emph{executor} is shown as $E_3$. 
Then, the green command is selected as anchor command because $N_1,N_3,N_4$ believe we should commit it first. 
After the commitment of green command, the executor should wait for \emph{log} sets submitted to push forward the ordering phase.
}
\label{fig:phalanx-process-consensus}
\end{figure*}

The executor module aims to get the final ordering (total ordering) according to the \emph{log} stream in $\mathrm{consensus}_i.\mathbb{S}$. 
To achieve \emph{anchor-based} ordering, we need to find the anchor commands with specific ordering rules. In this part, we would like to introduce the generation of final ordering.

\begin{table}[t]
    \centering
    \caption{\textbf{Structures Notations for Executor.}}
    \begin{tabular}{ l l }
        \toprule[1pt]
        Notation        & Description \\
        \midrule

        $\mathrm{executor}_i$ & The executor module for node $N_i$. \\
        \midrule
        
        $c$      & A command info to collect essential information for \\
                 & one command's commitment. \\
        \bottomrule[1pt]
    \end{tabular}
    \label{table:StructuresNotationExecutor}
\end{table}

\vspace{1mm} \noindent  \textbf{Structures}. 
The command $r$ is the basic unit for Byzantine ordered consensus. 
Here, as is shown in Table~\ref{table:StructuresNotationExecutor}, we take command info $c$ to collect essential information for command's commitment. 
We say that $c \to r$ if and only if $c$ is used to collect essential information for command $r$. 
The command info $c$ that $c \to r$ contains: 
(1) $d$, the digest of command that $d \gets r.d$.
(2) $O$, An $n$-sized vector of \emph{log}s where $O[i]$, initialized to $\bot$, tracks the \emph{log} $o$ from $N_i$ that $o \to r$, and the method $|c.O|$ would return the number of element in $O$ which is not $\bot$.
(3) $T$, a set that is used to store the timestamps. 
Whenever $o$ is added into $c.O$, $o.t$ would be added into $T$. 
The timestamps in $T$ are sorted in ascending numerical order.
If the length of $c.T$ is greater or equal to $2f+1$, then the $(f+1)$-th one would be recognized as trusted timestamp.

\begin{table}[t]
    \centering
    \caption{\textbf{States Notations for Executor.}}
    \begin{tabular}{ l l }
        \toprule[1pt]
        Notation        & Description \\
        \midrule
        $D$             & A set of strings, tracks the digests of commands \\
                        & which have been committed. \\
        \midrule

        $C$             & A vector of command info, where $C[d]$ refers \\
                        & to the command info $c$ whose digest is equal to $d$.\\
        \midrule
        
        $Q$             & An $n$-sized vector of FIFO queues, where $Q[j]$ is the \\
                        & FIFO queue $\mathbb{Q}_j$. \\
        \midrule
        $\mathbb{Q}_j$  & The queue used to record the \emph{log}s from $N_j$ \\
                        & according to the commitment order into \emph{executor}. \\
        \midrule
        
        $\theta_j$      & The front-\emph{log} in $\mathbb{Q}_j$ that $\theta_j \gets Q[j].read\_front()$. \\
                        & If $\mathbb{Q}_j.len()=0$, then $\theta_j \gets \bot$. \\
        \midrule
        
        $\Theta$        & An $n$-sized vector of front-\emph{log}s, where $\Theta[j]$ tracks \\ 
                        & the front-\emph{log} $\theta_j$. \\
        
        \bottomrule[1pt]
    \end{tabular}
    \label{table:StatesNotationExecutor}
\end{table}

\vspace{1mm} \noindent  \textbf{States}.
Each node $N_i$'s executor maintains the states as is shown in Table~\ref{table:StatesNotationExecutor}.
The set $D$ is used to track the committed commands. 
The vector $C$ is used to track the command info in executor module.
The $n$-sized vector $Q$ is used to store the commands submitted from consensus module. 
The $Q[j]$ refers to the FIFO queue $\mathbb{Q}_j$, which is used store the \emph{log}s of $N_j$. 
The $\theta_j$ refers to the front-\emph{log} in $\mathbb{Q}_j$. 
And we use an $n$-sized vector $\Theta$ to track the $\theta_j$.

\vspace{1mm} \noindent \textbf{Actions}. 
Whenever $\mathrm{consensus}_i.\mathbb{S}$ is not empty, take the front element in it that $S \gets \mathrm{consensus}_i.\mathbb{S}.front()$ and try to process $S$ with the ordering process as follows.

Step 1. 
Commit \emph{log}s in $S$ one by one into \emph{executor} module.
For each \emph{log} $o$ in $S$, read the command info $c$ from $C$ at first that $c \gets C[o.d_r]$ (if there doesn't exist such a command info, initiate it). Update $c.O[o.i]$ with $o$ and add $o$ into the end of FIFO queue $Q[o.i]$.

Step 2. Find the anchor-set. 
Initialize an empty anchor-set $F$. 
Read every FIFO queue in $Q$. While reading $\mathbb{Q}_i$, if the front-\emph{log} $\theta_i$ points to a committed command that $\theta_i.d_r \in D$, then remove it and read the latest front-\emph{log}. 
Repeat the process till the front-\emph{log} $\theta_i$ is $\bot$ or points to uncommitted command. Add each $\theta_i$ into a vector $\Theta$ that $\Theta[i]$ tracks the front-\emph{log} $\theta_i \gets Q[i].read\_front()$. 
If there exists commands $\{r\}$ that at least $f+1$ front-\emph{log}s point to $r$, then put all these command info $c \to r$ into $F$, return the anchor-set $F$. If not, jump into \emph{alter path}.

\emph{alter path}. For all the command info with $|c.O| \leq 2f+1$, get the command info $c$ with the lowest trusted timestamp. Then regard the command $c \to r$ as anchor command $r_a$. 
If $c \to r_a$, that we can find $c'$, $c' \to r'$, that there is not a \emph{reliable context} $r_a \prec r'$, then add $r'$ into $F$. 
Repeat this process until no such a $c'$ can be found or the $c'$ added into $F$ has $|c'.O|<2f+1$.

Step 3. Check the front set. 
For each command info $c$ in $F$, if $|c.O| < f+1$, then remove $c$ from $F$.
If $\exists c \in F$ that there are at least $f+1$ non-empty values in $c.O$ but $|c.O|$ is less than $2f+1$, then directly return an empty anchor-set.

Step 5. Commit the commands according to $F$. 
First of all, sort the command info \{c\} in anchor-set $F$ according to the in ascending order by trusted timestamp of $c$. 
If the trusted timestamps are the same, then sort the command info alphabetically.
After that, according to the sorted command info set $\{c\}$, we would commit the command $r$ that $c \to r$ one by one. To obtain the command $r$, we could read the mempool or request it from others.

As is shown in Fig~\ref{fig:phalanx-process-consensus}, it is the journey for non-faulty nodes to create total ordering after each participant has generated partial ordering in its mempool.


After the commitment of commands, each non-faulty node gets the same total ordering to execute them. 
After the execution of the command, nodes would feedback the results to proposers. 
If the proposer has received $f+1$ the same response for one command, accept the execution result for it.

\subsection{Proof}

\begin{theorem}[Validity]
    If a non-faulty node appends $r$ into its final ordering, then at least one non-faulty node has selected $r$ into its partial ordering.
\end{theorem}

\begin{IEEEproof}
    Before command $r$ has been committed into final ordering, the node should collect at least $2f+1$ logs from different participants pointing to it. The logs are related to nodes' partial ordering. As the adversary can only control up to $f$ nodes, $r$ has been selected at least one non-faulty node.
\end{IEEEproof}

\begin{theorem}[Consistency]
    For logs $o_1$ and $o_2$ from the same node, if their certificates are valid, then $o_1.n \neq o_2.n$.
\end{theorem}

\begin{IEEEproof}
    For logs $o_1$ and $o_2$ both generated by $N_i$, assume that they have valid certificates and $o_1.n=o_2.n=k$. 
    So that, $2f+1$ nodes have voted for $o_1$ and $o_2$ which means at least $f+1$ nodes have voted for logs from $N_i$ on sequence number $k$ twice. But the adversary can only control up to $f$ nodes. Therefore, here is a paradox.
\end{IEEEproof}

\begin{theorem}[Consistency]
    For non-faulty nodes $N_1$ and $N_2$, let $I_1$ and $I_2$ denote their total ordering. If there is context $r_1 \prec r_2$ in $I_1$, then we could eventually find $r_1 \prec r_2$ in $I_2$. 
\end{theorem}

\begin{IEEEproof}
    In consensus module, every non-faulty node will find the same log stream. While generating the total ordering, non-faulty node should process the log stream by order. Because of the strategy of SMR, if the log stream is the same, the state on each node would be the same. So that, the final ordering on each non-faulty node is the same.
\end{IEEEproof}

\begin{theorem}[Consistency]
    For nodes $\forall N_1\in N$ and $\forall N_2\in N$, regard their final order decision as $O_1$ and $O_2$. If $\exists c_1,c_2\in O_1 \land c_1\prec c_2$, then $\exists c_1,c_2\in O_2 \land c_1\prec c_2$.
\end{theorem}

\begin{theorem}[Finality]
    For valid verifiable logs $o_1$ and $o_2$ from the same node that $o_2.n=o_1.n+1$, if there is a committed order-batch $b$ that $o_2 \in b.H$, then each non-faulty node can receive the consistent $o_1$.
\end{theorem}

\begin{IEEEproof}
    For the verifiable logs $o_1$ and $o_2$ from $N_i$, $n_1 \gets o_1.n$, $n_2 \gets o_2.n$, $n_2 = n_1 + 1$. Because of the validation of $o_2$, $2f+1$ nodes have voted for it, which means they have already received $o_1$. While the commitment of logs in executor, we could request $o_1$ from others, since there are at most f malicious nodes, everyone could receive the valid $o_1$ at last.
\end{IEEEproof}

\begin{theorem}[Anchor Linearity]
    Let $R_a$ denote the set of anchor commands generated by Phalanx.  
    There is a partial ordering set $\langle R_a, \prec_r \rangle$.
\end{theorem}

\begin{IEEEproof}
    For $\forall r_1, r_2 \in R_a$ and $r_1$ is committed before $r_2$, make an assumption that there isn't reliable context that $r_1 \prec_r r_2$. 
    If $r_1$ is selected by normal path, then at least $f+1$ nodes believe we should commit $r_1$ at first, so that there is $r_1 \prec_r r_2$. 
    If $r_1$ is selected by alter path, then the commands $\{r'\}$ without $r_1 \prec_r r'$ are committed at the same anchor-set, so that for the uncommitted $r_2$, there is reliable context that $r_1 \prec_r r_2$.
\end{IEEEproof}

As for the commands belonging to the same anchor-set, the \emph{timestamp-based} strategy can be taken to decide their ordering. 
For there are at most $f$ faulty nodes, the $(f+1)$-th largest timestamp is most likely sent from a non-faulty node, when we have received at least $2f+1$ \emph{log}s pointing to specific command from different nodes.

\section{Implementation}

We implement Phalanx as a modular component in Go and should employ it with state-of-the-art BFT protocol. 
Bamboo\cite{gai2021dissecting} is an evaluation framework for chained-BFT protocols in Go. 
It has implemented multi kinds of chained-BFT protocols, e.g. chained HotStuff, StreamLet and etc.. 
Here, we complete Phalanx system on Bamboo chained-BFT implementations.
The following experiments concentrate on the Phalanx system employed on HotStuff in Bamboo. 
Refer to HotStuff as HS.
Refer to the extended protocol from HS as Phalanx-HS.

To amortize the cost of protocol, batching is a common optimization for performance improvement. 
In our implementation of Phalanx, we would like to make use of batching strategy. 
Each proposer could propose a command batch with $b$ requests in it.
Each node could generate one \emph{batched log} to declare its ordering preference for a series of commands that node could request ordering phase every $\Delta_o$ interval to declare its partial ordering in that duration. 
As for the performance, the batching strategy amortize the cost of hash calculation and signature verification, and Phalanx could achieve a more satisfying performance with batching strategy. 

\section{Experiment and Evaluation}

In this section, we evaluate the performance and reliability of Phalanx system which is employed in BFT protocol implementations on Bamboo. 
The main concern is about the impact of Phalanx component on latency and throughput in our system.
To analyze it, we compare Phalanx system with its baseline BFT protocols in both LAN and WAN environments.
Next, we concentrate on the fair ordering functionality of Phalanx. 
After that, we compare the robustness of \emph{anchor-based} Phalanx with \emph{timestamp-based} strategy.

\subsection{Performance Evaluation}

\noindent\textbf{Baseline and Metrics}.
We take HS as the baseline of our performance evaluation. 
We would like to compare the throughput and latency between HS and Phalanx-HS in both LAN and WAN environment. 
In this part, we concentrate on the performance loss caused by Phalanx component and some other factors which could affect performance. 
The latency has ignored the transmission delay of commands. 

\noindent\textbf{LAN Deployment and Performance}. 
We deployed our evaluation on 4 instances in the same datacenter to compare the performance of baseline and Phalanx. 
Each Phalanx node ran on Aliyun ECS c7 with 8vCPU (Intel Xeon 8369B) and 16GiB memory. 

For each datacenter, we use the default batch size $b=200$ and the ordering interval $\Delta_o = 50$ms. 
The results are shown in Table~\ref{table:phalanx-performance-lan}.
If there is only 1 proposer ($p=1$) to propose commands in Phalanx-HS, the throughput of it is almost the same as HS while the latency of Phalanx-HS is 15.3\% higher than HS. 
If we increase the number of proposers from 1 to 4, the throughput of Phalanx-HS increases in direct proportion to the number of proposers, while the latency of Phalanx-HS has barely changed. 
Then, we evaluate the performance of HS with $b=800$, and we found that its performance is almost the same as Phalanx-HS with $b=200$ and $p=4$.

\begin{table}[t]
    \centering
    \caption{\textbf{Performance of Phalanx in LAN.}}
    \begin{tabular}{l r r}
        \toprule[1pt]
    
        \bfseries  &\bfseries throughput (txs/s)&\bfseries latency (ms)\\
        \midrule
        
        HS(b=200) & 38,680 & 54.1 \\
        \midrule
        
        Phalanx-HS(b=200, p=1) & 35,956 & 62.4 \\
        Phalanx-HS(b=200, p=2) & 51,815 & 63.8 \\
        Phalanx-HS(b=200, p=3) & 72,197 & 62.3 \\
        Phalanx-HS(b=200, p=4) & 100,564 & 63.6 \\
        \midrule
        
        HS(b=800) & 119,499 & 57.4 \\
        \bottomrule[1pt]
    \end{tabular}
    \label{table:phalanx-performance-lan}
\end{table}

\noindent\textbf{WAN Deployment and Performance}. 
We deployed our evaluation on 4 geo-distributed nodes to compare the performance of baseline and Phalanx. 
Each Phalanx node ran on Aliyun ECS c5e with 8vCPU (Intel Xeon 8269CY) and 16GiB memory. 
Servers are located in Silicon Valley, Frankfurt, London, and Tokyo. 

For each geo-distributed node, we use the default batch size $b=200$ and the ordering interval $\Delta_o = 200$ms. 
The results are as shown in Table~\ref{phalanx-performance-wan}.
If there is only 1 proposer ($p=1$) to propose commands in Phalanx-HS, the throughput of it is almost the same as HS. 
However, the latency of Phalanx-HS is 5.57\% lower than HS, which is satisfying. 
It is mainly because that the main bottleneck of geo-distributed system is network communication and the \emph{order}-batch which only contains the latest logs of each participant has reduced the volume of HS consensus proposal. 
Besides, each part of Phalanx (mempool, consensus, and executor) are running concurrently, which can make full use of computing resources. 
Then, we increase the number of proposers from 1 to 4, and also find that the throughput of Phalanx-HS increases in direct proportion to the number of proposers with barely changed latency in geo-distributed environment. 

\begin{table}[t]
    \centering
    \caption{\textbf{Performance of Phalanx in WAN.}}
    \begin{tabular}{l r r}
        \toprule[1pt]
    
        \bfseries  &\bfseries throughput (txs/s)&\bfseries latency (ms)\\
        \midrule
        
        HS(b=200) & 1,288 & 1,324.6 \\
        \midrule
        
        Phalanx-HS(b=200, p=1) & 1,294 & 1,250.8 \\
        Phalanx-HS(b=200, p=2) & 2,515 & 1,264.8 \\
        Phalanx-HS(b=200, p=3) & 3,781 & 1,142.5 \\
        Phalanx-HS(b=200, p=4) & 5,196 & 1,252.1 \\
        \bottomrule[1pt]
    \end{tabular}
    \label{phalanx-performance-wan}
\end{table}

\noindent\textbf{Evaluation}. 
Compared to the baseline, Phalanx incurs signatures and more network communications to process verifiable \emph{log}s. 
However, the \emph{order}-batch reduces the volume of the state-of-the-art BFT proposals. 
Then, as for the latency for Phalanx, it is about 15\% higher than HS in LAN and could be slightly lower than HS in WAN environment. 
As for throughput, Phalanx is almost the same as HS if there is only 1 proposer, and the throughput of Phalanx increases in direct proportion to the number of proposers, while the latency of Phalanx-HS has barely changed. 
It can be considered that Phalanx does not bring too much performance loss, and it has advantages in some situations.

\subsection{Fault-Tolerance Functionality}

\noindent\textbf{Simulation for Ordering Manipulation.}
In this part, we simulate the ordering manipulation to verify the ability for Phalanx to resist attacks.
For the commands generated by the same proposer, there should be an intuitively distinct ordering that we need to commit the earlier proposed command at first. 
So that, we add sequence number on commands proposed to note the intuitive distinct ordering. 
To simulate the Byzantine behaviour, we inject some codes to make the adversary disorder the commands it has received and generate wrong partial ordering. 
With the monitoring on whether the commands are committed according to the sequence number, we can evaluate the ability to resist ordering manipulation. 

\begin{table}[htbp]
    \centering
    \caption{\textbf{Fault Tolerance on Ordering.}}
    \begin{tabular}{l r r}
        \toprule[1pt]
    
        \bfseries &\bfseries reordered commands ratio \\
        \midrule
        
        following byzantine & 99.8\%\\
        \midrule
        
        $f=0$ & 0\% \\
        $f=1$ & 0\% \\
        $f=2$ & 21.1\% \\ 
        \bottomrule[1pt]
    \end{tabular}
    \label{table:ordering-tolerance}
\end{table}

Table~\ref{table:ordering-tolerance} shows the change of reordered commands ratio with the number of Byzantine nodes. If there is no adversary that $f=0$, no intuitively distinct context is broken. 
If there is one Byzantine node and follow the attacker's ordering, then almost all the intuitively distinct contexts are destroyed. 
However, after the activation of Phalanx strategy, only almost every intuitively distinct context has been resisted. 
If the number of adversarial nodes is larger than the fault-tolerance threshold, we can find there are part of command pairs with intuitively distinct context have been reordered.

\subsection{Adversary Resistance}

In this part, we compare the adversarial attack resistance ability between \emph{timestamp-based} strategy and Phalanx. 
For Byzantine fault tolerance protocols, if the number of Byzantine nodes is no larger than the fault-tolerance threshold, we expect the intuitively distinct contexts should be preserved. 
Here, we do not implement a \emph{timestamp-based} protocol\cite{zhang2020byzantine}\cite{kursawe2020wendy}. 
We compare the Phalanx with \emph{timestamp-based} strategy directly.

\begin{table}[htbp]
    \centering
    \caption{\textbf{Adversarial Attack Resistance Comparison.}}
    \begin{tabular}{l c c}
        \toprule[1pt]
    
        \bfseries $f$ &\bfseries Phalanx &\bfseries Timestamp-based \\
        \midrule
        
        0 & \checkmark & \checkmark \\
        \midrule
        
        1 & \checkmark & \checkmark \\
        2 & \checkmark & \checkmark \\
        3 & \checkmark & $\times$ \\
        4 & \checkmark & $\times$ \\
        5 & \checkmark & $\times$ \\
        \bottomrule[1pt]
    \end{tabular}
    \label{table:tolerance-comparison}
\end{table}

We deployed our evaluation on 16 nodes.
Each Phalanx node ran on Aliyun ECS c7 with 8vCPU (Intel Xeon 8369B) and 16GiB memory. 
We inject codes to make some of them malicious and set 2 proposers.
Increase the byzantine node number from 0 to threshold ($f=5$) and detect the ratio of commands which have been reordered. 
Here, let us consider that the system has resisted manipulation attack if the ratio of reordered commands is less than 0.5\%. 
As is shown in Table~\ref{table:tolerance-comparison}, both Phalanx and \emph{timestamp-based} strategy could reserve the intuitively distinct context without any Byzantine nodes. 
However, if there exists Byzantine nodes in current situation, the \emph{timestamp-based} strategy sometimes cannot prevent manipulation attack, while Phalanx could still reverse intuitively distinct contexts.

\begin{figure}[ht]
\centering
\includegraphics[scale=0.5]{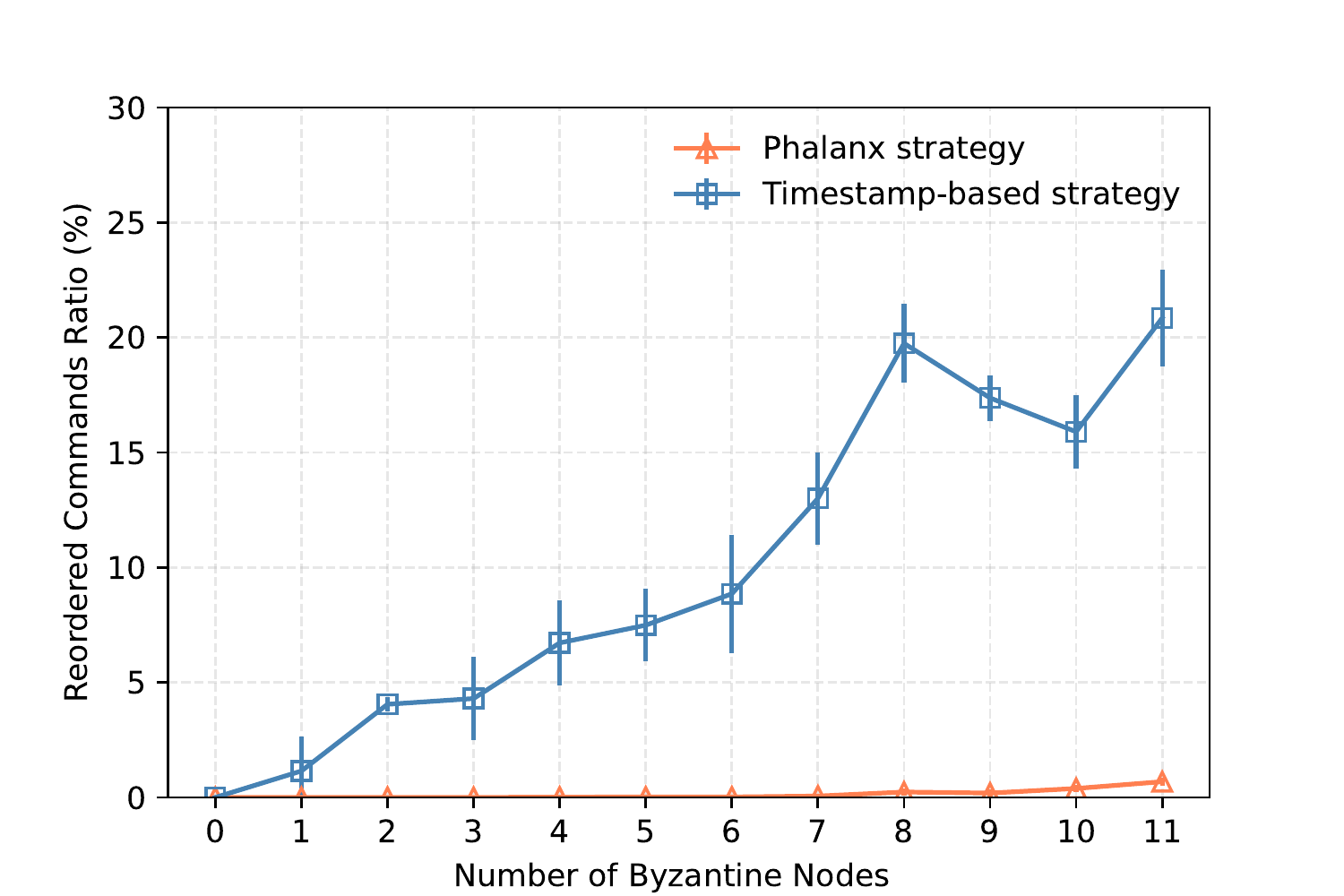}
\caption{Results for experiment on adversary resistance comparison between Phalanx and \emph{timestamp-based} strategy in cluster with 34 nodes.}
\label{fig:adversary-comparison}
\end{figure}

To take further evaluation on adversary resistance, we deployed our evaluation on 34 nodes.
Each Phalanx node ran on Aliyun ECS c7 with 4vCPU (Intel Xeon 8369B) and 8GiB memory. 
We also inject codes to make some of them malicious and set 2 proposers.
Increase the byzantine node number from 0 to threshold ($f=11$) and detect the reordered commands ratio. 
As is shown in Fig~\ref{fig:adversary-comparison}, the probability of \emph{timestamp-based} strategy causing command reordered which has destroyed intuitively distinct context is significantly higher than Phalanx. 
As the number of Byzantine nodes increases within the threshold, the commands ordered by \emph{timestamp-based} strategy will be much easier to be attacked, while the reordered commands ratio through Phalanx does not increase rapidly. 
We found that the Phalanx performs better in resisting ordering manipulation than \emph{timestamp-based} strategy.

Besides, we find that, whenever reordered commands ratio of Phalanx goes up, there are more anchor-sets selected through \emph{alter path}. 
It is an attention-worth problem that can be further studied in the future.

\section{Related Works} \label{sec:related}


    
        

Aequitas\cite{kelkar2020order} proposed by Kelkar et al. in 2020 are the first class of protocols which have achieved weak \emph{order-fairness} in both synchronous and asynchronous situations. 
To deal with Condorcet Paradox, Aequitas construct a relation graph of transactions and find out all of the strongly connected components (SCCs) and the transactions in the same SCC would constitute Condorcet Paradox. 
After that Aequitas could find deterministic ordering to execute each SCC, and the transactions in the same SCC would be applied in the same batch. 
It could be considered as a \emph{batch-based} ordering protocol. 
In 2021, Themis\cite{kelkar2021themis} was proposed by Kelkar et al.. 
It is the upgraded \emph{batch-based} ordering protocol from Aequitas.
It has resolved the liveness problem and has reduced communication complexity.
However, to detect SCCs in relation graph, the commonly used algorithm such as Tarjan\cite{tarjan1972depth} and Kosaraju\cite{sharir1981strong} have high computation complexity which is related to the number of transactions and their context. It is an obvious system bottleneck. 
The experiment of Themis shows that its throughput is about 20\% of HS with the same batch-size in LAN.

Pompe\cite{zhang2020byzantine} proposed by Zhang et al. has also proposed a practical protocol to deal with ordering manipulation.
Here, transactions are ordered according to the medium timestamp that it is a naturally linear ordering indicator which could avoid Condorcet Paradox. 
With this simplified ordering strategy, Pompe has achieved higher throughput at competitive latencies compared with the state-of-the-art BFT protocols. 
As the indicator is constructed by timestamps, the ordering strategy in Pompe could be regarded as a \emph{timestamp-based} ordering protocol. 
In the same year, Wendy\cite{kursawe2020wendy} proposed by Kursawe et al. has also achieved \emph{order-fairness} with medium timestamp similar to Pompe. 

Although, without detecting SCCs, \emph{timestamp-based} ordering strategy seems like to have resolved the bottleneck of Byzantine ordered protocols, it just provides each node an opportunity to decide the final order, which reduces the impact of the adversary but cannot prevent arbitrary behavior of manipulating ordering. 
As is shown in Table~\ref{table:medium-time-fail}. 
There are 4 nodes that $N_3$ is a malicious node. 
Here is a pair of commands $c_1$ and $c_2$ with intuitively distinct context $c_1 \prec c_2$ that they are proposed by the same proposer and $c_1$ is proposed several seconds before $c_2$. 
$N_2$ and $N_4$ receive these commands slightly later, while $N_1$ received these commands earlier. 
If we select $N_1$, $N_2$, and $N_3$ as the final set, then $c_1$'s medium timestamp 2 would be smaller than $c_1$ that such pair of commands would be reserved. 

\begin{table}[htbp]
    \centering
    \caption{\textbf{Medium timestamp failure.}}
    \begin{tabular}{l c c c c}
        \toprule[1pt]
        \bfseries &\bfseries $N_1$  &\bfseries $N_2$ &\bfseries $N_3$ &\bfseries $N_4$\\ 
        \midrule
        
        \bfseries $c_1$ &  0 &  3 &  3 &  3\\ 
        \bfseries $c_2$ &  1 &  4 &  2 &  4\\ 
        \bottomrule[1pt]
    \end{tabular}
    \label{table:medium-time-fail}
\end{table}

Fiary\cite{teeorder}, proposed by Stathakopoulou et al. is an ordering system with the assistance of TEE to prevent front-running attacks. 
Fairy needs to introduce specific hardware to support the operation of the protocol, and its correctness depends on the reliability of the TEE itself.
So that, the scope of its application is relatively limited.
In 2021, Cachin et al. also theoretically proposed a quick order fairness atomic broadcasting protocol \cite{cachin2021quick}, which can ensure that messages are delivered in different fair orders. However, it has not been implemented. 

\section{Conclusion}

To propose an efficient Byzantine ordered consensus protocol, we propose \emph{anchor-based} ordering strategy and design a protocol called Phalanx based on it.
Phalanx is a practical Byzantine ordered protocol, which can be employed in each state-of-the-art BFT protocol.
The ordering phase in Phalanx does not take too much performance loss that only 15\% more latency and slightly decrease of throughput in LAN. Because of the optimization of communication, the latency of Phalanx is lower than HS in WAN. 
As for resisting ordering manipulation, Phalanx is better than the \emph{timestamp-based} strategy. 
Besides, the ratio of anchor-sets generated through \emph{alter path} seems to be taken to track the reliability of Phalanx, which could be further studied in the future.

\bibliographystyle{IEEEtran}
\bibliography{ref}

\end{document}